\documentclass{article}
\usepackage{spconf,amsmath,graphicx}
\usepackage{cite}[sorted]
\usepackage{multirow}

\usepackage{amsmath}
\usepackage{etoolbox}
\usepackage{bm}
\usepackage{amsthm}
\usepackage{mathtools}
\usepackage{algorithm}
\usepackage{algpseudocode}
\usepackage{xspace}

\newcommand{\MyMapTemplateNoPrefix}[3]{\expandafter#1\csname#3\endcsname{#2{#3}}}
\forcsvlist{\MyMapTemplateNoPrefix {\def} {\mathbf} } {0,1,a,b,c,d,e, f, g, h, i, j, k, l, m, n, o, p, q, r, u, v, w, x, y, z} 

\newcommand{\MyMapTemplatePrefix}[4]{\expandafter#1\csname#3#4\endcsname{#2{#4}}} 
\forcsvlist{\MyMapTemplatePrefix {\def} {\mathcal}{c}} {A,B,C,D,E,F,G,H,I,J,K,L,M,N,O,P,Q,R,S,T,U,V,W,X,Y,Z}  
\forcsvlist{\MyMapTemplatePrefix {\def} {\mathbf} {b}} {A,B,C,D,E,F,G,H,I,J,K,L,M,N,O,P,Q,R,S,T,U,V,W,X,Y,Z} 
\forcsvlist{\MyMapTemplatePrefix {\DeclareMathOperator} {} {} } {trace, argmax, argmin, MixUp, Shuffle, Concat, aggregate, features, Freeze, Unfreeze, mAP, Melt}

\def\x{{\mathbf x}}

\title{HalluAudio: Hallucinate frequency as concepts for few-shot audio classification}
\name{Zhongjie Yu, Shuyang Wang, Lin Chen, Zhongwei Cheng}
\address{Wyze Labs, Inc.}
\begin{document}
%
\maketitle
\begin{abstract}
Few-shot audio classification is an emerging topic that attracts more and more attention from the research community. Most existing work ignores the specificity of the form of the audio spectrogram and focuses largely on the embedding space borrowed from image tasks, while in this work, we aim to take advantage of this special audio format and propose a new method by hallucinating high-frequency and low-frequency parts as structured concepts. Extensive experiments on ESC-50 and our curated balanced Kaggle18 dataset show the proposed method outperforms the baseline by a notable margin.
The way that our method hallucinates high-frequency and low-frequency parts also enables its interpretability and opens up new potentials for the few-shot audio classification.
\end{abstract}

\section{Introduction}
\label{sec:intro}

Deep learning has shown extraordinary performance in recognizing and discriminating different sounds in recent years. However, such good performance relies on a large amount of high-quality labeled data. Although few-shot learning has been proposed to learn robust classifiers from only a few examples, most existing works only apply to the image classification tasks~\cite{matchingnetwork,prototypical,relationnet,maml,imprinting,closerlook,activation,TPN}. Collecting a large amount of labeled image data is time-consuming and expensive, whereas collecting audio data annotations is even more difficult. For example, it is intuitive for humans to label an image with ``dog'' by looking at the entire image with a glimpse; however, it usually takes a much longer time to annotate audio with ``dog barking'' as it takes more efforts to listen and understand the entire audio clip. Furthermore, it is almost impossible for humans to annotate an audio clip by only looking at its spectrogram. Additionally, humans rely more heavily on visual cues than audio cues, therefore it is sometimes difficult to give precise labels by only listening to audio clips such as the classic confusion between ``baby crying'' and ``cat meowing''~\cite{meowingcry}. All the above-mentioned challenges impose a great demand for few-shot audio classification algorithms.

\begin{figure}[t]
    \centering
    \includegraphics[width=0.45\textwidth]{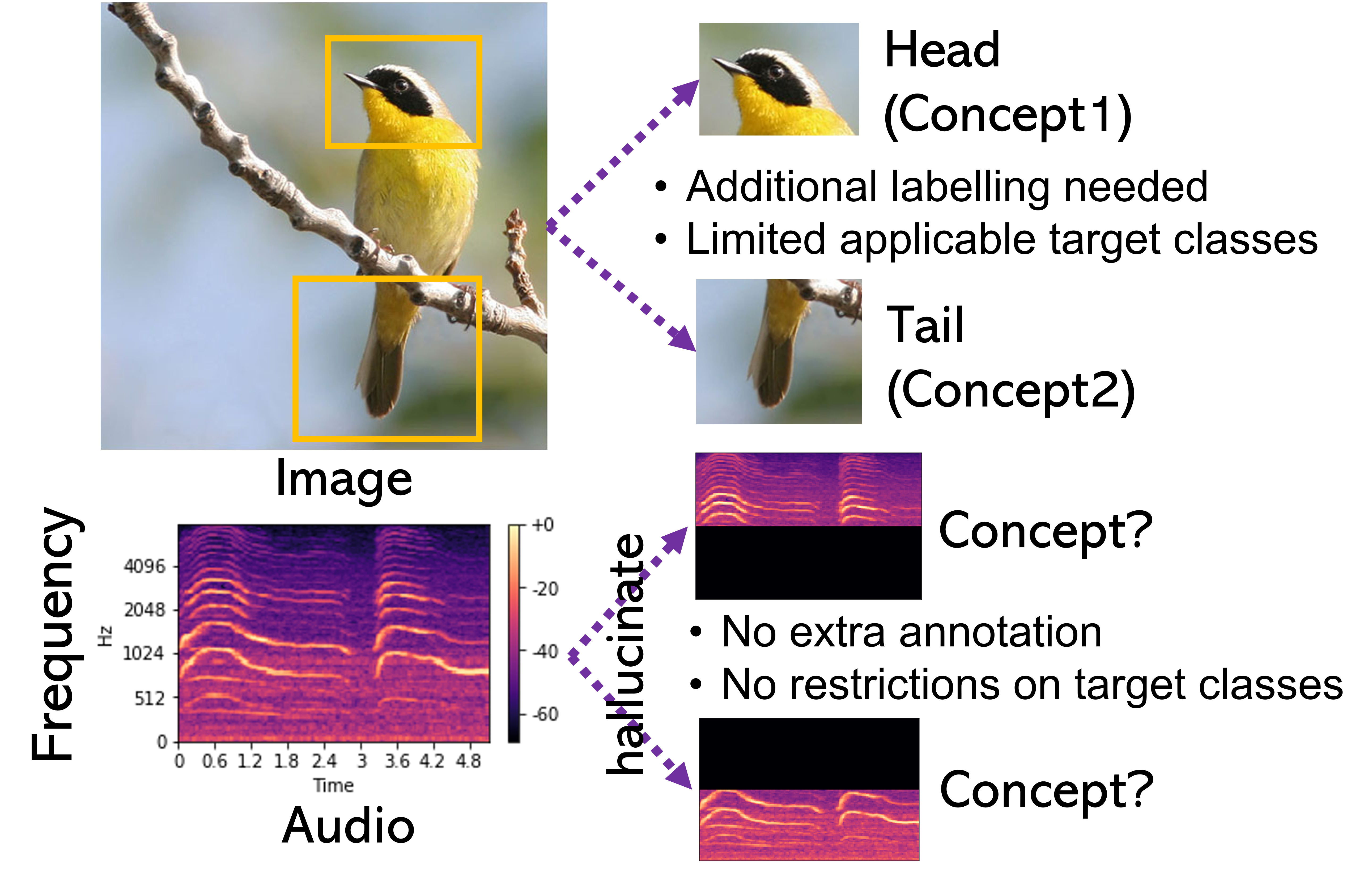}
    \vspace{-0.1in}
    \caption{Illustration of our HalluAudio idea. Detailed structure information in images is utilized as ``concepts'' to improve few-shot learning performance, but it lacks effectiveness due to the additional labeling cost and the restricted scope of ``concepts''. However, for audio data, the frequency domain in the spectrogram is discriminative. We hallucinate the high-frequency and low-frequency areas as hidden concepts lying in the spectrogram, which are utilized to improve the few-shot audio classification.}
    \label{fig:illustration}
    \vspace{-0.1in}
\end{figure}


However, there is only a handful of work addressing few-shot audio classification~\cite{fsactraining,fsacstudy,fsaed,metaaudio}. Among those, most works attempted to directly apply general few-shot learning methods like Prototypical Network~\cite{prototypical}, MAML~\cite{maml} on audio data. Beyond that, a very limited number of works tried to develop new methods for few-shot audio classification. For example, \cite{audiognn} proposed an attentional GNN for audio classification, \cite{attentionsimi} developed an attention similarity module and \cite{raresoundeventdetection} integrated CTM~\cite{ctm}, TPN~\cite{TPN} and MixUp~\cite{mixup} with audio data augmentation to build a task-adaptive module. Nonetheless, all these methods are still focusing only on the extracted unstructured embedding space rather than the audio spectrogram itself, just like the most common way for few-shot image classification. In other words, those methods could be reasonable for handling visual images but may not be capable of highlighting the special modality of the audio spectrogram in the image format.

In terms of images themselves rather than their embeddings, \cite{conceptlearner} is the first meta-learning work to dig out the utility of concepts from those. What are {\it concepts} in images? They are some items with structured knowledge such as the head, wing, and tail of a bird. Given those human-interpretable concepts, \cite{conceptlearner} is able to improve the performance in a straightforward way as well as by introducing reasoning for the recognition, which is different from other methods only targeting the unstructured embedding space that is prone to be a black box.

Although audio spectrograms can be presented in the same format as visual images and fed into similar neural networks, it is unclear whether it is possible to use interpretable concepts for audio spectrograms. First and foremost, does it exist the ``real'' structured concepts in audio spectrograms that can be cognized by humans? We can easily recognize the head, wings or tail of birds, but it remains unexplored whether there is a similar pattern for the audio spectrogram. Secondly, a strong prerequisite of using those interpretable concepts is that samples belonging to similar classes should share that structured knowledge. For example, sparrows and terns both have heads, wings and tails, whereas laptops have neither of those concepts, so there is a barrier to using concepts when classifying laptops and sparrows. Lastly, annotating the bounding boxes and labels for the concepts in images needs a large amount of extra workloads. Consequently, this is a notable restriction to apply the utilization of structured concepts. Only very limited numbers of datasets provide those extra detailed labels. For example, the CUB dataset~\cite{cub2011} has the detailed locations of 15 concepts in each image, without which it's not feasible to learn those structured concepts.

Motivated by those challenges in audio spectrograms, we propose HalluAudio, a meta-learning method that hallucinates high-frequency and low-frequency parts in the audio spectrogram as structured concepts and then utilizes those concepts to build frequency-specific learners. More specifically, the high-frequency prototype and low-frequency prototype are constructed from the high-frequency part and low-frequency part in the spectrogram, respectively. Then HalluAudio aggregates high-frequency, low-frequency, and original prototypes as the representation for a given audio spectrogram.  With this way of ``hallucinating'' audio spectrogram, the previous mentioned challenges are addressed as: (1) it provides a practical way of depicting concepts for audio spectrogram; (2) it does not rely on the assumption that samples should belong to similar classes because every audio spectrogram can have concepts in the high and low-frequency areas; (3) it needs no extra labeling work, because all the high and low-frequency areas can be derived from some specific ranges of Hz in the spectrogram. To the best of our knowledge, this is the first method directly exploring the pattern in the audio spectrogram for few-shot audio classification, which is essentially different to methods leveraging the unstructured embedding space.

\section{Proposed Method}

\begin{figure}[t]
    \centering
    \includegraphics[width=0.5\textwidth]{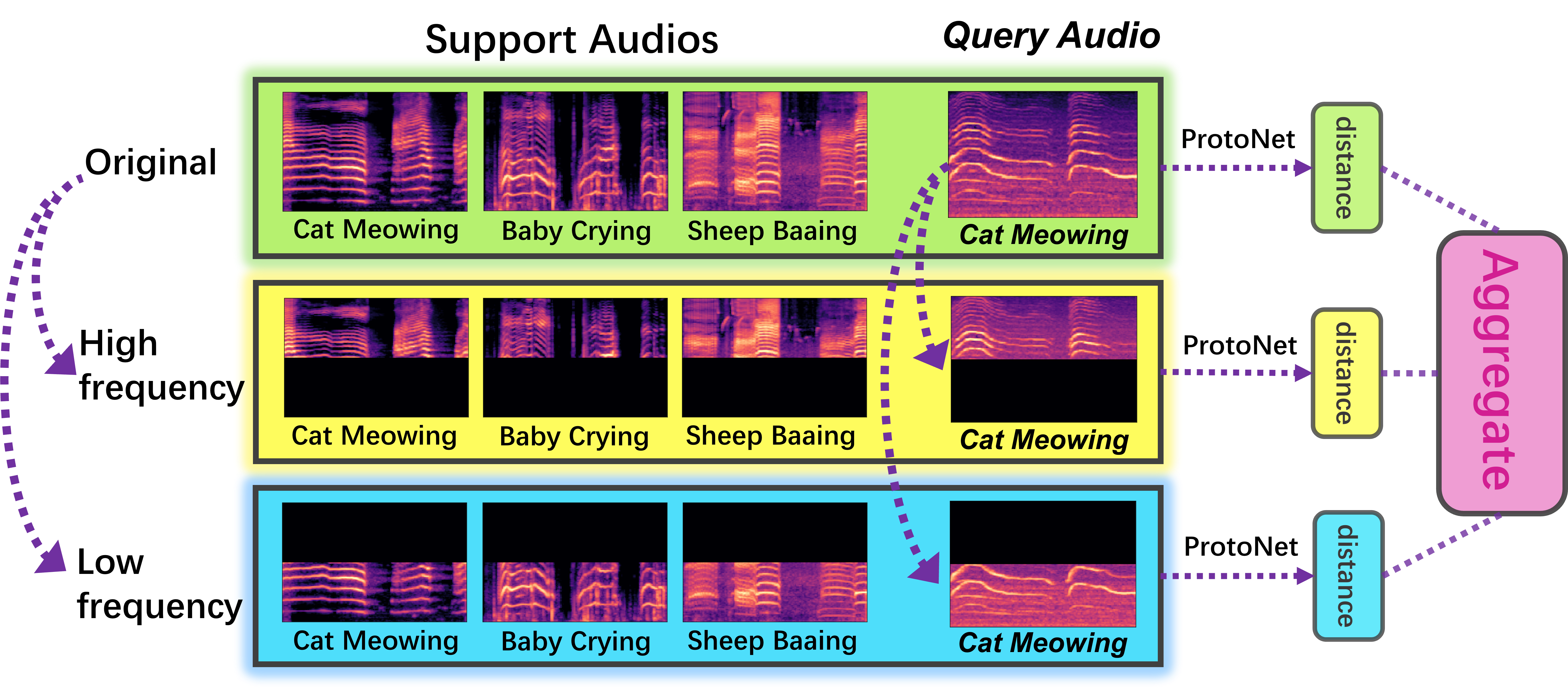}
    \vspace{-0.1 in}
    \caption{The proposed HalluAudio in a 3-way 1-shot setting. We compute three types of distance and aggregate them for a final decision with the prototypical network. The first type is from the query audio's embedding with the prototypes of support audios. The second and third types of distance are computed from the embedding of the query audio in the high-frequency and low-frequency domain with the corresponding prototypes of support audios, respectively. High-frequency and low-frequency domains in the spectrogram are served as two hallucinated concepts. }
    \label{fig:flowchart}
    \vspace{-0.1 in}
\end{figure}

\subsection{Problem Definition}
In few-shot audio classification, the goal is to train a robust classifier for novel-class audio data given only a few examples. During training, we are given a training dataset $\cD_{base}$ containing many-shot audios from base classes $\cC_{base}$. During testing, we are given a test dataset $\cD_{novel}$ containing few-shot audios from novel classes $\cC_{novel}$ where $\cC_{base} \cap \cC_{novel}=\emptyset$. In an $N$-way $K$-shot task, we are given a support set $\cS=\{(\cX_s,\cY_s)\}$ and one query sample $\x_q$ \cite{attentionsimi}, where $\cX_s$ consists of $N\times K$ audios $(\x_1,\x_2,\dots,\x_{N\times K})$, $\cY_s$ are their class labels $(y_1,y_2,\dots,y_{N\times K})$ and $\x_q$ belongs to those $N$ classes. 

\subsection{Method Description} 
As summarized in Figure \ref{fig:flowchart}, the input of the neural networks for a given audio file's waveform $\w_l$ is the log mel spectrogram $\x_l$, which represents the amplitude of the audio in $T$ $\times$ $F$ dimension where $T$ is the time-domain range and $F$ is the frequency range in mel-scale. In our HalluAudio, we hallucinate frequency from different ranges as the structured concepts embedded in the log mel spectrogram. More specifically, we denote the frequency hallucination group as $\cM=\{\m^{(n)}\}_{n=1}^N$, where $\m^{n}$ is the $n$-th binary vector masking the frequency area, and $N$ is the number of masks. With this, the $n$-th frequency prototype for class $k$ is $\p_k^{(n)}=\frac{1}{|\cS_k|}\sum_{(\x_l,y_l)\in\cS_k}{f^{(n)}(\x_l\cdot \m^{(n)})}$, where $f^{(n)}(\cdot)$ is the feature extractor for $n$-th hallucination. Then, the final probability of a given $\x$ belonging to class $k$ combines the results from the original spectrogram and different frequency groups:
$$\frac{\exp \left( -d\left(f(\x),\p_k\right)-\sum_n{d\left(f^{(n)}(\x\cdot \m^{(n)}),\p_k^{(n)}\right)} \right )}{\sum_{k}\exp \left(-d\left(f(\x),\p_k\right)-\sum_n{d\left(f^{(n)}(\x\cdot \m^{(n)}),\p_k^{(n)}\right)} \right )},$$
where $ f(\cdot)$ is the feature extractor for the whole spectrogram which has the same structure with $f^{(n)}(\cdot)$, $\p_k$ is the prototype of the whole spectrogram $\p_k=\frac{1}{|\cS_k|}\sum_{(\x_l,y_l)\in\cS_k}{f(\x_l)}$, and $d(\cdot)$ is the Euclidean Distance. 

\noindent{\bf Remark:} We point out that a similar operation of setting time and frequency areas to zero introduced in SpecAug~\cite{specaugment} is only a way of data augmentation, which aims to add noises to the log mel spectrogram to improve robustness. Our method significantly differs from this augmentation way on the whole idea and motivation.





\section{Experiments}

In this section, we evaluate our proposed HalluAudio and conduct an ablation study on the widely adopted ESC-50 dataset and our curated dataset from Kaggle18 for few-shot audio classification.

\subsection{Dataset Configuration}

The current research for few-shot audio classification lacks common agreements on criteria from dataset choices, processing, and evaluation metrics. To be consistent with \cite{attentionsimi} and fit the current research focus on fixed-length data, we choose ESC-50 dataset~\cite{esc50} which contains 50 classes and 40 samples per class with a fixed length of 5 seconds. In addition, we curate a balanced fixed-length dataset from Kaggle18 dataset which is originally variable-length data of 11,073 audios from 41 classes of Audioset Ontology~\cite{audioset}. 

All audio samples from ESC-50 and Kaggle18 datasets are down-sampled from 44.1kHz to 16kHz. We extract log mel spectrogram using $librosa$~\cite{librosa}. The number of Mel bands is set to 128. The highest frequency is set to 8000Hz. 
The hop size is set to 502 for ESC-50 and 201 for Kaggle18 to generate spectrograms with 160$\times$128 dimensions. The power spectrogram is converted to decibel units. Because there are not enough details for generating the log mel spectrogram for ESC-50 in \cite{attentionsimi}, our generated log mel spectrogram could be slightly different from their provided files using their codes.

\subsection{Training and Evaluation}
With the very limited public codes in this domain, we strictly follow the training pipeline used by \cite{attentionsimi}. Note that the episode-building strategy in \cite{attentionsimi} is slightly different from the one commonly used in few-shot image classification. During the testing stage, each sample is served as a query and the corresponding N-way K-shot supports are randomly sampled from the rest of the test data to build an episode. To get more reliable results and confidence intervals, we conduct the sampling 50 times instead of only once as in \cite{attentionsimi}.

The network backbone is used the same as \cite{attentionsimi}, which is also adopted in \cite{knowledge}. This backbone consists of 3 blocks. Each block is composed of a 3$\times$3 convolutional layer, batch normalization, ReLU layer, and 4$\times$4 max pooling layer. The initial learning rate is set to $0.01$ and SGD is used for optimization with the weight-decay set to 0.0001. For ESC-50, we use the same strategy as \cite{attentionsimi} in which the learning rate is divided by 10 after every 20 epochs. For Kaggle18, the learning rate is divided by 10 after every 30 epochs. Both are trained for 60 epochs.

\begin{table*}[t]
    \centering
    \begin{tabular}{c|c|c|c|c|c}
    \hline
    Dataset &    Method  & 5-way 1-shot & 5-way 5-shot & 10-way 1-shot & 10-way 5-shot \\
    \hline
       \multirow{2}*{ESC-50} & Baseline \cite{attentionsimi} & 69.77 $\pm$ 0.62 & 83.47 $\pm$ 0.48 & 
        54.51 $\pm$ 0.66&  
        71.36 $\pm$ 0.56\\
      ~ &  HalluAudio & {\bf 71.88 $\pm$ 0.60} & {\bf 86.46 $\pm$ 0.46} &
        {\bf 57.12 $\pm$ 0.64}&
        {\bf 75.20 $\pm$ 0.58}\\
        \hline
        \multirow{2}*{Kaggle18} & Baseline \cite{attentionsimi} & 57.58 $\pm$ 0.63 &
          70.69 $\pm$ 0.55&
         43.67 $\pm$ 0.62  &58.12 $\pm$ 0.58 \\
       ~& HalluAudio &{\bf 59.35 $\pm$ 0.65} & {\bf 73.92 $\pm$ 0.57} & {\bf 44.50 $\pm$ 0.64} & {\bf 61.80 $\pm$ 0.61}\\
       \hline
        
    \end{tabular}
    \caption{Accuracy (in \%) on ESC-50 and Kaggle18 datasets with 95\% confidence interval. Note    Baseline result is not identical to \cite{attentionsimi} because of the parameters for the log mel spectrogram and testing sampling times as mentioned in section 3.2.}
    \label{tab:mainresult}
\end{table*}

\begin{table*}[tp]
    \centering
    \begin{tabular}{c|c|c|c|c|c}
    \hline
    Dataset &    Method  & 5-way 1-shot & 5-way 5-shot & 10-way 1-shot & 10-way 5-shot \\
    \hline
       \multirow{4}*{ESC-50} & Baseline \cite{attentionsimi} & 69.77 $\pm$ 0.62 & 83.47 $\pm$ 0.48 & 
        54.51 $\pm$ 0.66&  
        71.36 $\pm$ 0.56\\
      ~ &  Time Concept &  70.66 $\pm$ 0.61 
        & 82.89 $\pm$ 0.48
        &55.67 $\pm$ 0.67
        &70.60 $\pm$ 0.53 \\
        ~ &Gain(time) & 0.89 & -0.58 & 1.16 & -0.76\\
        ~ & {\bf Gain(freq.)} & {\bf 2.11}  & {\bf 2.99} & {\bf 2.61} & {\bf 3.84} \\
        \hline
        \multirow{4}*{Kaggle18} & Baseline \cite{attentionsimi} & 57.58 $\pm$ 0.63 & 
          70.69 $\pm$ 0.55&
         43.67 $\pm$ 0.62  &58.12 $\pm$ 0.58 \\
       ~& Time Concept & 58.13 $\pm$ 0.65 & 71.30 $\pm$ 0.57& 43.95 $\pm$ 0.61 & 58.77 $\pm$ 0.60\\
           ~ &Gain(time) & 0.55 & 0.61 & 0.28 & 0.65\\
        ~ & {\bf Gain(freq.)} & {\bf 1.77}  & {\bf 3.23} & {\bf 0.83} & {\bf 3.68} \\
       \hline

    \end{tabular}
    \caption{Ablation study of hallucinating concepts in time domain vs. frequency domain in the spectrogram. Taking concepts in the time domain with the same network does not notably improve the performance and it even harms the results in some cases.
    }
    \label{tab:timedomain}
    \vspace{-0.1 in}
\end{table*}

\begin{figure}
    \centering
    \includegraphics[width=0.3\textwidth]{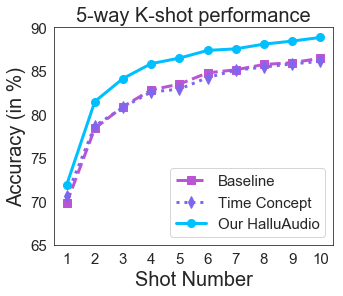}
    \vspace{-0.1 in}
    \caption{HalluAudio (Frequency Concept) vs. Time Concept vs. Baseline in 5-way K-shot settings.}
    \label{fig:threeline}
    \vspace{-0.2 in}
\end{figure}

\subsection{Experimental Results}

Table \ref{tab:mainresult} shows the results and confidence intervals for the baseline (prototypical network) and our proposed HalluAudio. It is clearly showing that our method outperforms the baseline by a large margin. For ESC-50, the gains are 2.11\%, 2.99\%, 2.61\%, 3.84\% for 5-way 1-shot, 5-way 5-shot, 10-way 1-shot, 10-way 5-shot, respectively. For Kaggle18 dataset, the gains are 1.77\%, 3.23\%, 0.83\%, 3.68\%, respectively. 

To validate that the gain is from hallucinating high frequency and low frequency as concepts rather than additional weights in the network, we conduct an ablation study of hallucinating time domain as concepts. In particular, we hallucinate the first half of the time as one concept, and the second half of the time as another concept. Notably, the network constructed using this way of hallucinating ``time'' concepts has the same weights as the network using frequency concepts.  As shown in Table \ref{tab:timedomain}, for ESC-50, although there is a little improvement for 1-shot, the time concepts make a negative contribution for 5-shot. This reflects the human intuition that audio with uncertain patterns has little structured information in the time domain. On Kaggle18, hallucination from the time domain improves the performance a little bit because we curate fixed-length audios around the peak of the waveform. In this case, the first half of the spectrogram in the time domain could stand for the starting period of the audio and the second half stands for the ending period. However, the significantly inferior performance by hallucinating time concepts compared with hallucinating frequency concepts strongly show the rationality of our method. To have a more comprehensive comparison, we add the results of three methods in 5-way K-shot settings for ESC-50 in Figure \ref{fig:threeline}.

\subsection{Frequency Importance}
To better show the reasoning behind the hallucination of frequency areas, we calculate the frequency importance in some representative classes. Specifically, we select 5 representative classes and their 5-way 5-shot episodes. Given a query, we classify it by the distance only from (1) its high-frequency embedding and support samples' high-frequency prototypes; (2) its low-frequency embedding and support samples' low-frequency prototypes. In this way, we get two correctly classified numbers of queries in all episodes: $Q_{high}$ and $Q_{low}$.

\begin{figure}[]
    \centering
    \includegraphics[width=0.38\textwidth]{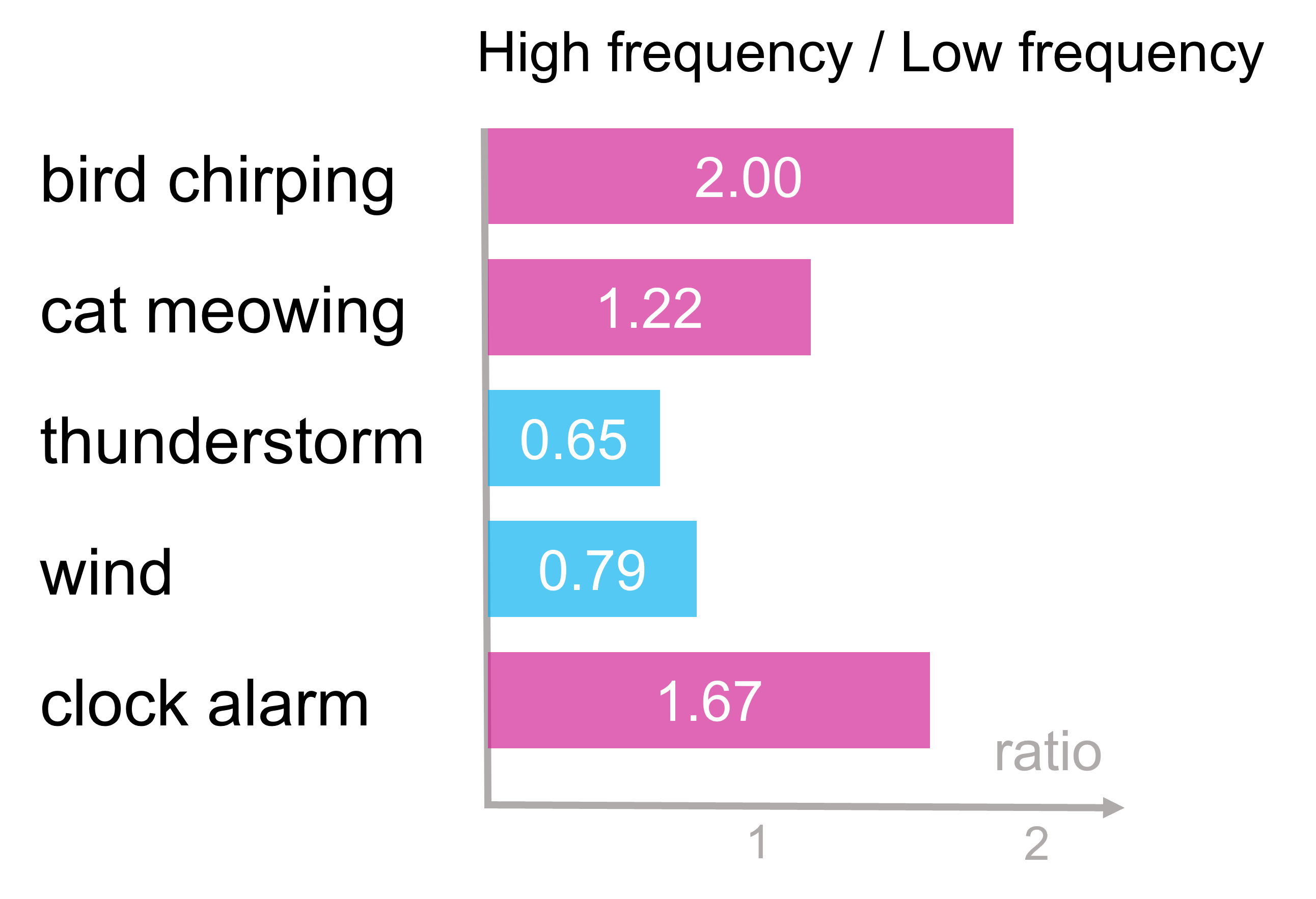}
    \vspace{-0.1in}
    \caption{The frequency importance in representative classes.}
    \label{fig:ratio}
    \vspace{-0.1in}
\end{figure}

\begin{figure}[]
    \centering    \includegraphics[width=0.4\textwidth]{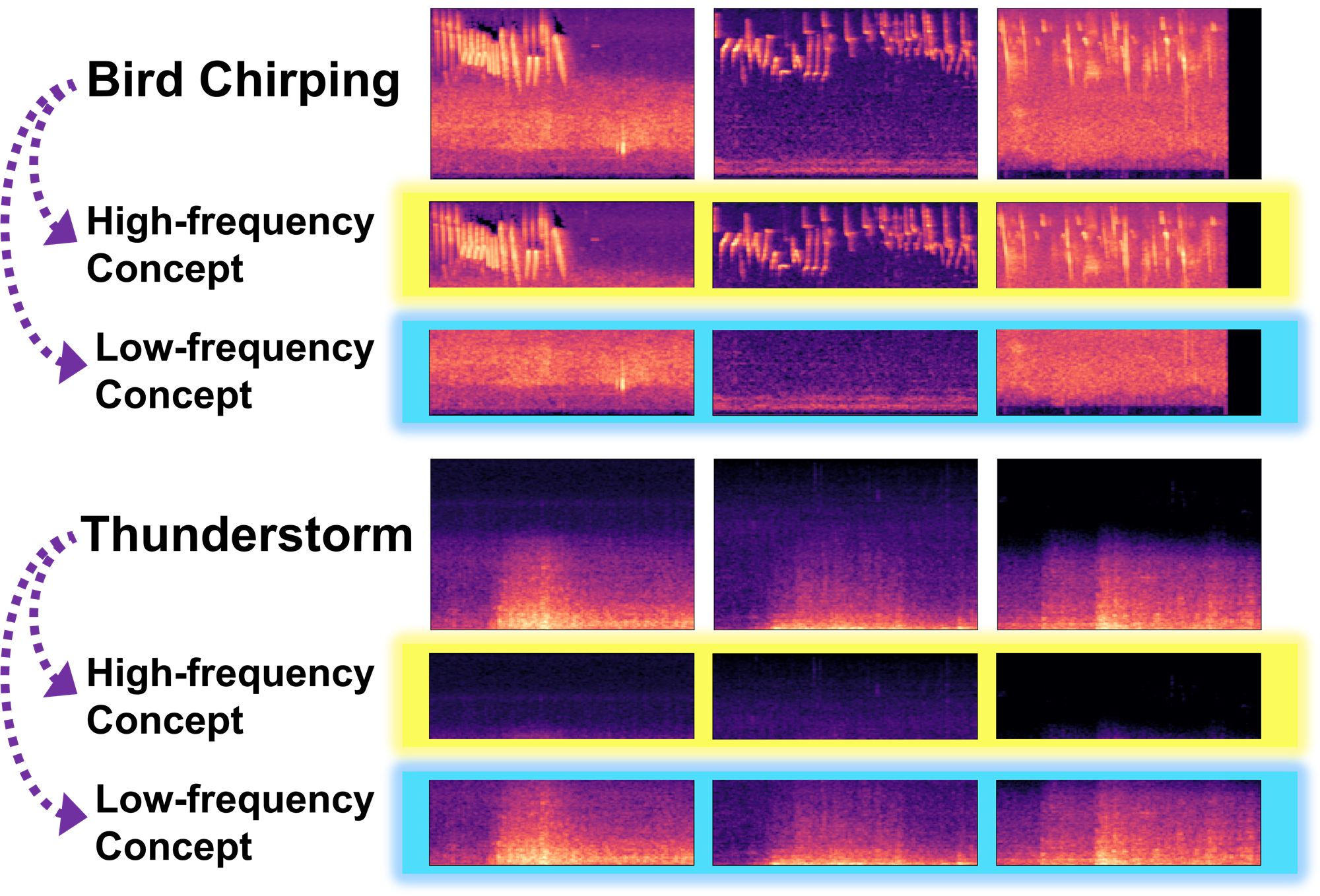}
        \vspace{-0.1in}
    \caption{Illustration of high/low-frequency concepts for bird-chirping and thunderstorm. Bird-chirping has more similar patterns in high-frequency concepts, whereas
thunderstorm is mostly depicted by low-frequency concepts.}
    \label{fig:birdthun}
    \vspace{-0.1in}
\end{figure}

Given the number, we calculate the frequency importance by $\frac{Q_{high}}{Q_{low}}$.
A ratio greater than 1 means the high frequency has more importance and a ratio less than 1 means the low frequency is more important. As shown in Figure \ref{fig:ratio}, the ratio matches common sense: birds' chirping has more information in the high-frequency area, whereas thunderstorm is more depicted in the low-frequency area. Furthermore, we also show some examples in Figure \ref{fig:birdthun} which matches our analysis.

\section{Conclusion}
We have proposed a simple yet effective method for few-shot audio classification. Our method, dubbed as HalluAudio, hallucinates high-frequency and low-frequency areas in the spectrogram as structured concepts. Compared with the real concepts used for few-shot image classification, hallucinating concepts take the advantage of the special format of spectrogram and do not need any extra labeling work or have any restrictions on specific classes. Extensive experiments on ESC-50 and Kaggle18 datasets demonstrate the rationality of our proposed solution. To the best of our knowledge, this is the first work focusing on and utilizing the specificity of audio spectrogram format with interpretability in few-shot audio classification and opens a new horizon in this area.

\vfill\pagebreak

\bibliographystyle{IEEEbib}
\bibliography{fsac}

\begin{thebibliography}{10}

\bibitem{matchingnetwork}
Oriol Vinyals, Charles Blundell, Timothy Lillicrap, Daan Wierstra, et~al.,
\newblock ``Matching networks for one shot learning,''
\newblock in {\em NIPS}, 2016, pp. 3630--3638.

\bibitem{prototypical}
Jake Snell, Kevin Swersky, and Richard Zemel,
\newblock ``Prototypical networks for few-shot learning,''
\newblock in {\em NIPS}, 2017, pp. 4077--4087.

\bibitem{relationnet}
Flood Sung, Yongxin Yang, Li~Zhang, Tao Xiang, Philip~HS Torr, and Timothy~M
  Hospedales,
\newblock ``Learning to compare: Relation network for few-shot learning,''
\newblock in {\em CVPR}, 2018, pp. 1199--1208.

\bibitem{maml}
Chelsea Finn, Pieter Abbeel, and Sergey Levine,
\newblock ``Model-agnostic meta-learning for fast adaptation of deep
  networks,''
\newblock in {\em ICLR}, 2017.

\bibitem{imprinting}
Hang Qi, Matthew Brown, and David~G. Lowe,
\newblock ``Low-shot learning with imprinted weights,''
\newblock in {\em CVPR}, 2018.

\bibitem{closerlook}
Wei-Yu Chen, Yen-Cheng Liu, Zsolt Kira, Yu-Chiang Wang, and Jia-Bin Huang,
\newblock ``A closer look at few-shot classification,''
\newblock in {\em ICLR}, 2019.

\bibitem{activation}
Siyuan Qiao, Chenxi Liu, Wei Shen, and Alan~L Yuille,
\newblock ``Few-shot image recognition by predicting parameters from
  activations,''
\newblock in {\em CVPR}, 2018.

\bibitem{TPN}
Yanbin Liu, Juho Lee, Minseop Park, Saehoon Kim, Eunho Yang, Sung~Ju Hwang, and
  Yi~Yang,
\newblock ``Learning to propagate labels: Transductive propagation network for
  few-shot learning,''
\newblock {\em arXiv preprint arXiv:1805.10002}, 2018.

\bibitem{meowingcry}
Karen McComb, Anna~M Taylor, Christian Wilson, and Benjamin~D Charlton,
\newblock ``The cry embedded within the purr,''
\newblock {\em Current Biology}, vol. 19, no. 13, pp. R507--R508, 2009.

\bibitem{fsactraining}
Jordi Pons, Joan Serr{\`a}, and Xavier Serra,
\newblock ``Training neural audio classifiers with few data,''
\newblock in {\em ICASSP}, 2019, pp. 16--20.

\bibitem{fsacstudy}
Piper Wolters, Chris Careaga, Brian Hutchinson, and Lauren Phillips,
\newblock ``A study of few-shot audio classification,''
\newblock {\em arXiv preprint arXiv:2012.01573}, 2020.

\bibitem{fsaed}
Bowen Shi, Ming Sun, Krishna~C Puvvada, Chieh-Chi Kao, Spyros Matsoukas, and
  Chao Wang,
\newblock ``Few-shot acoustic event detection via meta learning,''
\newblock in {\em ICASSP}. IEEE, 2020, pp. 76--80.

\bibitem{metaaudio}
Calum Heggan, Sam Budgett, Timothy Hospedales, and Mehrdad Yaghoobi,
\newblock ``Metaaudio: A few-shot audio classification benchmark,''
\newblock {\em arXiv preprint arXiv:2204.02121}, 2022.

\bibitem{audiognn}
Shilei Zhang, Yong Qin, Kewei Sun, and Yonghua Lin,
\newblock ``Few-shot audio classification with attentional graph neural
  networks.,''
\newblock in {\em Interspeech}, 2019, pp. 3649--3653.

\bibitem{attentionsimi}
Szu-Yu Chou, Kai-Hsiang Cheng, Jyh-Shing~Roger Jang, and Yi-Hsuan Yang,
\newblock ``Learning to match transient sound events using attentional
  similarity for few-shot sound recognition,''
\newblock in {\em ICASSP}. IEEE, 2019, pp. 26--30.

\bibitem{raresoundeventdetection}
Chendong Zhao, Jianzong Wang, Leilai Li, Xiaoyang Qu, and Jing Xiao,
\newblock ``Adaptive few-shot learning algorithm for rare sound event
  detection,''
\newblock {\em arXiv preprint arXiv:2205.11738}, 2022.

\bibitem{ctm}
Hongyang Li, David Eigen, Samuel Dodge, Matthew Zeiler, and Xiaogang Wang,
\newblock ``Finding task-relevant features for few-shot learning by category
  traversal,''
\newblock in {\em CVPR}, 2019, pp. 1--10.

\bibitem{mixup}
Hongyi Zhang, Moustapha Cisse, Yann~N Dauphin, and David Lopez-Paz,
\newblock ``mixup: Beyond empirical risk minimization,''
\newblock {\em arXiv preprint arXiv:1710.09412}, 2017.

\bibitem{conceptlearner}
Kaidi Cao, Maria Brbic, and Jure Leskovec,
\newblock ``Concept learners for few-shot learning,''
\newblock in {\em ICLR}, 2020.

\bibitem{cub2011}
C.~Wah, S.~Branson, P.~Welinder, P.~Perona, and S.~Belongie,
\newblock ``{The Caltech-UCSD Birds-200-2011 Dataset},''
\newblock Tech. {R}ep. CNS-TR-2011-001, California Institute of Technology,
  2011.

\bibitem{specaugment}
Daniel~S Park, William Chan, Yu~Zhang, Chung-Cheng Chiu, Barret Zoph, Ekin~D
  Cubuk, and Quoc~V Le,
\newblock ``Specaugment: A simple data augmentation method for automatic speech
  recognition,''
\newblock in {\em Interspeech}, 2019, pp. 2613--2617.

\bibitem{esc50}
Karol~J Piczak,
\newblock ``Esc: Dataset for environmental sound classification,''
\newblock in {\em ACM international conference on Multimedia}, 2015, pp.
  1015--1018.

\bibitem{audioset}
Jort~F Gemmeke, Daniel~PW Ellis, Dylan Freedman, Aren Jansen, Wade Lawrence,
  R~Channing Moore, Manoj Plakal, and Marvin Ritter,
\newblock ``Audio set: An ontology and human-labeled dataset for audio
  events,''
\newblock in {\em ICASSP}. IEEE, 2017, pp. 776--780.

\bibitem{librosa}
Brian McFee, Colin Raffel, Dawen Liang, Daniel~P Ellis, Matt McVicar, Eric
  Battenberg, and Oriol Nieto,
\newblock ``librosa: Audio and music signal analysis in python,''
\newblock in {\em Proceedings of the 14th python in science conference}.
  Citeseer, 2015, vol.~8, pp. 18--25.

\bibitem{knowledge}
Anurag Kumar, Maksim Khadkevich, and Christian F{\"u}gen,
\newblock ``Knowledge transfer from weakly labeled audio using convolutional
  neural network for sound events and scenes,''
\newblock in {\em ICASSP}. IEEE, 2018, pp. 326--330.

\end{thebibliography}

\end{document}